# Twist-Reconfigurable van der Waals Moiré Photonic Crystals

**Hugo Quard[1,2,†*], Jiyun Kim[1,2*], Anastasiia Zalogina[1,2], Xuerong Hu[3], Evan Williams[1,2], Oscar J. Palma Chaundler[3], Owen R. Wolley[3], Alexander Tartakovskii[3], Haoning Tang[4] and Igor Aharonovich[1,2,†]**

[1] School of Mathematical and Physical Sciences, University of Technology Sydney, Ultimo, New South Wales 2007, Australia

[2] ARC center of Excellence for Transformative Meta-Optical Systems, University of Technology Sydney, Ultimo, New South Wales 2007, Australia

[3] School of Mathematical and Physical Sciences, University of Sheffield, Sheffield, S3 7RH, United Kingdom

[4] School of Engineering and Applied Sciences, Harvard University, Cambridge, MA, USA

[*] These authors contributed equally to this work

[†] Corresponding to igor.aharonovich@uts.edu.au, hugo.quard@uts.edu.au

## Abstract

*Moiré photonics has emerged as a fascinating concept to design and in situ control of the optical bands. Moiré enabled light localisation arises from the relative twist between periodic layers, rather than from fixed, pre-fabricated cavity features. So far, however, the realisation of practical moiré photonic crystals in the visible range has been elusive, due to challenges in engineering nanoscale structures and twisting them dynamically post fabrication. Here, we realise a mechanically reconfigurable moiré photonic crystal, comprising from two patterned van der Waals crystals (tungsten di sulphide, $WS_2$) slabs separated by an optically active hexagonal boron nitride (hBN) spacer. We reconfigured the same pair of $WS_2$ slabs from a twist angle of 3.8° to 8.4° and reconstructed their three-dimensional dispersion using momentum-resolved reflectivity spectroscopy. Further, by reducing the twist angle between the slabs, we observe a denser manifold of folded and hybridised resonances that coincides with a 30-fold enhancement of emission from embedded colour centres. Our results open exciting opportunities for in-situ dispersion engineering and programmable light matter interactions employing van der Waals nanostructures.*

## INTRODUCTION

Moiré patterns are among the most familiar and established optical illusions, and yet have been seldomly employed for deterministically controlling and manipulating light at the nanoscale (***1*–*5***). When two photonic lattices (or periodic structures) are rotated relative to one another, their reciprocal lattices become mismatched, opening new momentum-transfer channels and folding the parent dispersions into a reduced moiré Brillouin zone (***1*, *6*, *7***). Interlayer coupling hybridises these folded modes into twist-dependent photonic minibands (***8*, *9***).

The idea that sharp moiré resonances can substantially reduce the group velocity of light was introduced in fibre gratings more than two decades ago (***3***). However, its extension to two-dimensional nanophotonic structures has become experimentally accessible only recently through advances in high-resolution nanofabrication, and momentum-resolved spectroscopy.
In this regime, Bloch-like modes retain the periodic structure of the parent lattice but acquire a slowly varying moiré envelope that localises them near selected stacking configurations. This often occurs at the maximally overlapping AA regions, while suppressing out-of-plane radiation through interference (***6–8*, *10*, *11***). By orchestrating guided resonances and collective Bloch modes, individual photonic-crystal (PhC) slabs have become a powerful platform for controlling spontaneous emission, nonlinear interactions and the coupling of light to free-space radiation (***12–17***).
Consequently, implementing moiré photonics by combining two PhC slabs introduces a reconfigurable degree of freedom absent in monolithic moiré photonic structures: the relative twist between the constituent PhCs. Unlike structures in which the moiré pattern is fixed during fabrication (***18–22***), the twist angle can be adjusted after assembly, enabling continuous control over the resulting photonic states. Varying the twist simultaneously reshapes the optical dispersion, localisation, and radiation properties without modifying the constituent PhCs. The novelty of the moiré cavity is therefore not the existence of cavity arrays itself, but the emergence of localised, high Q, small-mode-volume states, and their associated miniband manifolds, from a continuously adjustable geometry. This form of light manipulation has already enabled magic-angle and collective lasing (***19*, *21*, *23***), cavity quantum electrodynamics and emitter-lifetime control (***24*, *25***), nonlinear frequency conversion (***26*, *27***) and twist-dependent beam steering (***28*, *29***). Despite this progress, realizing reconfigurable PhCs, particularly in the visible range, remains challenging, as it requires two high-quality patterned slabs, precise control over their nanoscale separation, and a gap small enough for the eigenmodes of the two lattices to overlap and hybridize.
Here, we realise a mechanically reconfigurable moiré PhC assembled from two patterned tungsten di sulphide ($WS_2$) slabs separated by a hexagonal boron nitride (hBN) spacer. These van der Waals (vdW) crystals offer a compelling photonic platform: transition-metal dichalcogenides can combine a high refractive index with strong excitonic and nonlinear optical responses (***30–32***), while hBN can host optically active quantum emitters (***33***). Their atomically smooth, dangling-bond-free surfaces and weak interlayer adhesion also enable independently fabricated photonic layers to be assembled through dry transfer and, under suitable conditions, restacked at a different twist angle. Recent $WS_2$ bilayer metasurfaces have also demonstrated unidirectional guided resonances associated with Dirac-band continua (***34***), highlighting the versatility of vdW materials for photonic band engineering.
Fig. 1 illustrates the concept of twist-induced moiré PhCs in a vdW platform. Each PhC slab is fabricated from a $WS_2$ flake patterned with a hexagonal lattice of air holes with lattice constant $a$. Two such slabs are separated by a pristine hBN spacer, as shown in Fig. 1a and b. The high refractive index of $WS_2$ (n: ~4.1 (***35*, *36***)) supports strongly confined guided resonances, whereas the hBN layer defines the inter-slab separation $d$, maintains optical contrast between the slabs, and provides an active host for quantum emitters. In this architecture, the twist angle becomes an additional structural degree of freedom, allowing a single PhC design to be reconfigured into different moiré geometries through rotational stacking. In real space, a relative twist between the two PhC slabs produces a moiré superlattice with a period $L_m = a/2\sin(\theta/2)$ as illustrated in Fig. 1b. Reducing the twist angle therefore increases the size of the moiré unit cell and extends the spatial modulation imposed by the superlattice. The inter-slab separation d provides a second independent parameter that controls the strength of the electromagnetic coupling between the two PhC slabs. Together, the twist angle and inter-slab separation define a mechanically reconfigurable moiré geometry, providing independent control over band folding and inter-slab hybridization. The corresponding moiré Brillouin zone is shown in Fig. 1c. It is constructed from the primitive reciprocal lattice vectors $b_1$ and $b_2$, defined through $a_i \cdot b_j = 2\pi\delta_{ij}$ where $a_i$ and $b_j$ are the primitive real- and reciprocal-space lattice vectors, respectively. Relative twisting reduces the size of the Brillouin zone and introduces the moiré reciprocal wavevector $g_m$. As illustrated in Fig. 1d, the moiré superlattice folds photonic bands from the constituent slabs into multiple reduced Brillouin zones, where inter-slab coupling hybridizes the folded bands and opens gaps at their crossings. This process generates additional photonic bands and increases the density of optical states.

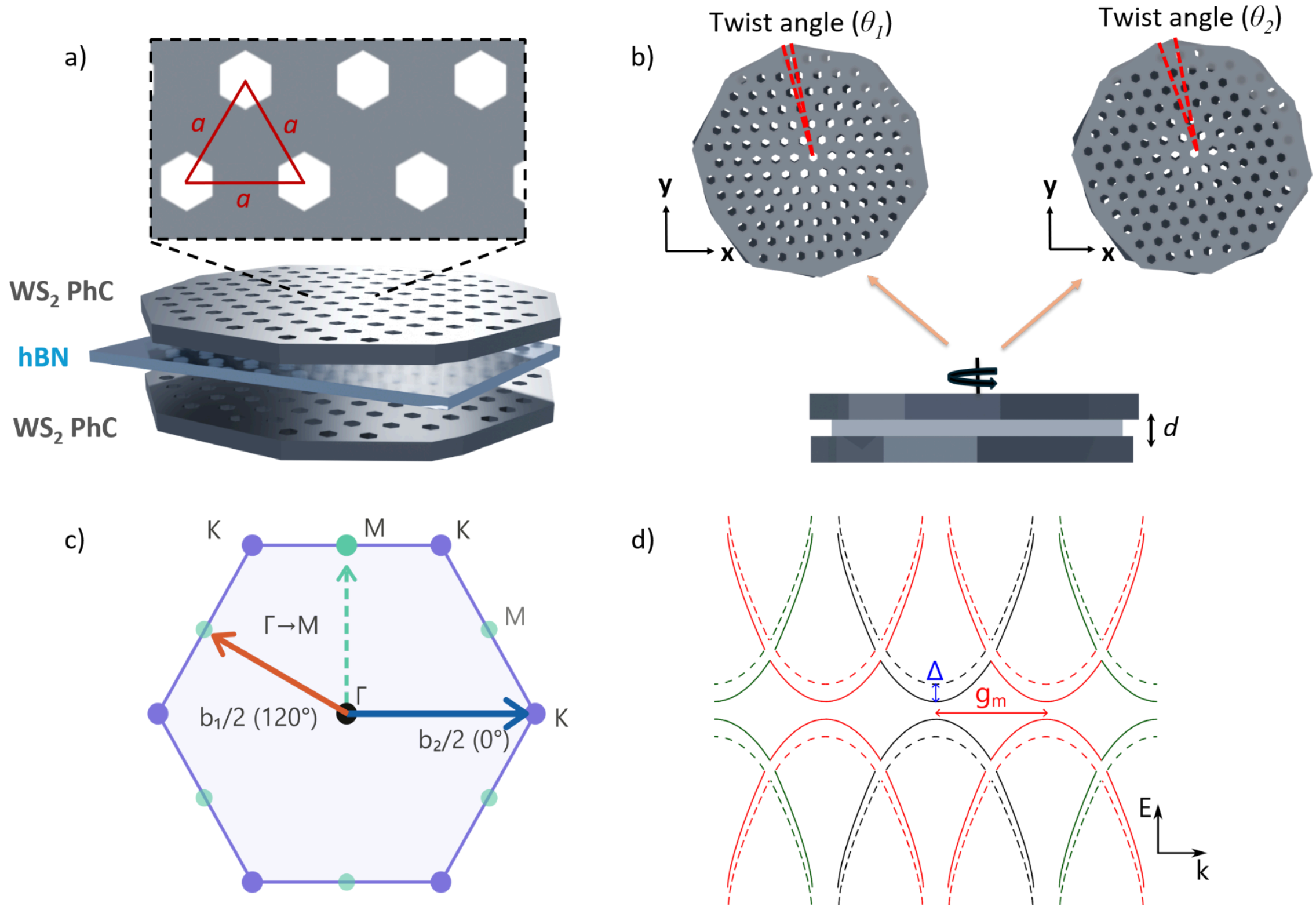

***Figure 1. Schematic of twisted-induced moiré in vdW PhC platform.*** *a) Schematic of the moiré heterostructure consisting of two fabricated $WS_2$ PhC slabs separated by a pristine hBN spacer. Each $WS_2$ PhC has identical hexagonal air holes with a lattice constant (a) of* 426 *nm. b) Schematic of side view of the moiré heterostructure separated by d (hBN thickness) and real-space views of the moiré superlattice formed at two representative twist*

*angles* $\theta_1$ and $\theta_2$ *with the red dotted line indicating the degree of twist of top PhC slab. c) Photonic Crystal Brillouin zone showing the reciprocal lattice vectors* $b_1/2$ *and* $b_2/2$*, the high-symmetry points* Γ*, M and K, and the momentum-space directions probed by angle-resolved spectroscopy. The vertical and horizontal directions correspond to the* $\Gamma \rightarrow M$ *and* $\Gamma \rightarrow K$ *paths, respectively. d) Schematic illustration of moiré band formation in momentum space. The black solid and dashed curves represent photonic bands derived from the two constituent PhC slabs and separated by a frequency offset* Δ *determined by the inter-slab separation. Folding by the moiré reciprocal wavevector* $g_m$ *generates the replicated bands shown in red and green. Inter-slab coupling hybridizes the folded bands at their crossings and opens gaps, resulting in additional photonic bands and an increased density of optical states.*

**RESULTS**

To investigate moiré band folding and hybridization by twist-induced moiré superlattices in vdW PhC, two $WS_2$ PhC slabs were assembled by polymer-assisted dry transfer, as shown in Fig. 2a. The top $WS_2$ PhC slab was first engaged and picked up, followed by the hBN flake. Then, the bottom slab is rotated to corresponding twist angle before the top PhC/hBN is released onto the bottom PhC slab. The process is repeated to dynamically reconfigure the twist angle between the top and the bottom PhCs, as will be shown below. In the current work, we elucidate moiré bands of the assembled bilayer PhC slabs at two twist angles: $\theta_1$: 3.8 ° and $\theta_2$: 8.4 ° (The twist angles are analysed in Fig. S2). Note that we first assembled it at $\theta_1$ and then twisted the same stack at $\theta_2$ after characterizing the moiré band at $\theta_1$.

The optical microscopy image of the final moiré heterostructure at $\theta_2$ is demonstrated in Fig. 2b. The periodicity and size of hexagonal holes in both top and bottom $WS_2$ PhC slab was further carried out by scanning electron microscopy (SEM), as shown in the inset of Fig. 2b. It is found that both slabs have a lattice constant of $a$ = 426 nm and hexagonal holes with a diameter of ~ 250 nm, which is used for further computational modelling.

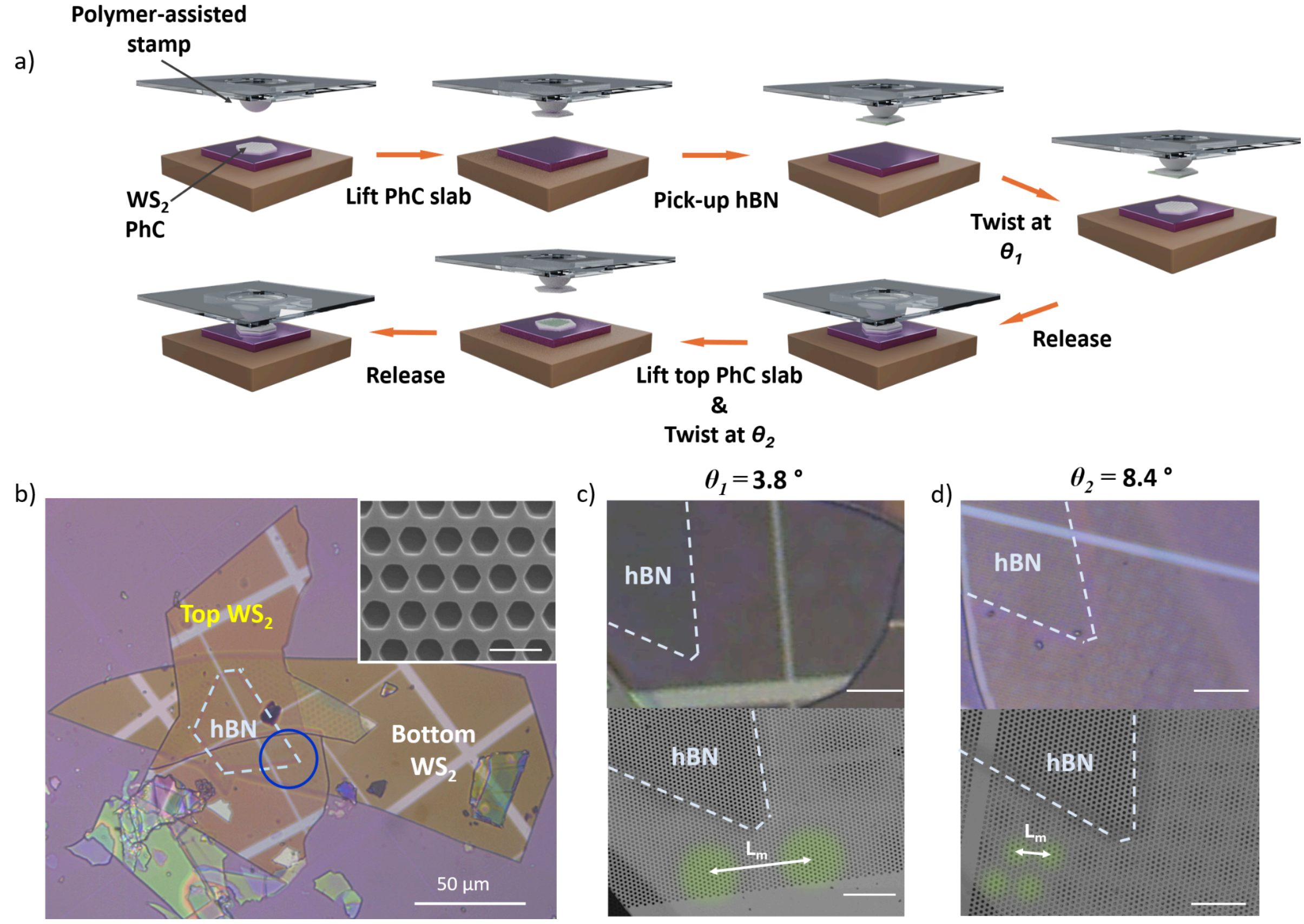

***Figure 2. Transfer process and moiré superlattices at two representative angles $\theta_1$* and *$\theta_2$*.** *a) Illustration of the workflow of the polymer-assisted dry transfer process. b) Optical microscopy image of the final moiré heterostructure twisted at $\theta_2$, representing the top and bottom $WS_2$ PhC slabs with the hBN spacer (marked as dashed line) with a scale bar of 50 µm. The inset is the SEM image of the bottom PhC slab with a scale bar of 500 nm, confirming the hexagonal holes with the lattice constant, a, of 426 nm. (c-d) is the SEM image (bottom) and optical image (top) corresponding to the designated region (marked as blue circle in (b)) at $\theta_1$ and $\theta_2$, respectively, representing the moiré superlattices (green circles in SEM images) with a moiré period ($L_m$) at $\theta_1$ and $\theta_2$.*

The designated region (marked as blue circle) in Fig. 2b was further investigated to confirm the moiré period at $\theta_1$ and $\theta_2$. As the twist angle decreases, the moiré period increases according to $L_m \approx a/\theta$ in the small-angle limit, where $\theta$ is expressed in radians. The optical contrast within the moiré superlattice can be seen in Fig. 2c and d for twist angles of $\theta_1$ and $\theta_2$. High-resolution SEM images (bottom) were further performed to verify the moiré period at $\theta_1$ and $\theta_2$. Twisted bilayer slabs at $\theta_1$ exhibit a moiré period ($L_m$) roughly 2.5-fold larger than that at $\theta_2$. The measured moiré period is slightly larger than the theoretical moiré period (2.2-fold larger) due to the divergent sensitivity of the moiré period in the small-angle regime.

The optical bands supported by both single-slab PhCs and moiré PhCs were then characterized using angle-resolved reflectivity measurements. These measurements are performed at room temperature using a back-focal-plane (BFP) imaging configuration, commonly referred to as k-space spectroscopy (*37*). A schematic of the experimental setup is shown in Fig. 3a, and the operating principle is described in detail in the Methods section. A collimated broadband beam is directed into the back aperture of a microscope objective, which focuses the light onto the sample. In this configuration, each incidence angle is associated with a specific in-plane momentum, and the reflected light is collected through the same objective. Owing to the Fourier-transform property of the objective, the sample plane and the objective BFP correspond to real-space and angular-space information, respectively. The measurement yields a three-dimensional reflectivity dataset $R(\theta_x, \theta_y, \lambda)$, where each spectrum is acquired for a fixed value of $\theta_x$ and the full dataset is reconstructed by scanning $\theta_x$. Throughout this work, the optical bands are represented in angular coordinates $(\theta_x, \theta_y)$, which are directly related to the in-plane wavevector through $k_{x,y} = k_0 / \sin(\theta_{x,y})$, where $k_0 = 2\pi/\lambda$.

The measured angle-resolved reflectivity spectra of the constituent PhC slabs are shown in Fig. 3b and c. Although both slabs share the same in-plane geometry, including the lattice constant and hole pattern, they possess markedly different thicknesses of 252 nm and 112 nm for the bottom and top slabs, respectively (Fig. S3). As a consequence, the two structures exhibit significantly different band dispersions. Unlike the case of two identical PhCs, where coupling occurs between similar photonic bands, the present architecture enables hybridization between spectrally and spatially distinct modes. This modal asymmetry provides an additional degree of freedom for engineering the optical response of the coupled system, as the interaction between dissimilar photonic modes can give rise to new hybridized states in the moiré structure. To establish a reference in the absence of moiré effects, Fig. 3d shows the arithmetic average of the spectra from the two individual slabs. This averaged band structure is used throughout the following analysis to distinguish genuinely moiré-induced photonic bands from features that originate from the independent responses of the individual slabs.

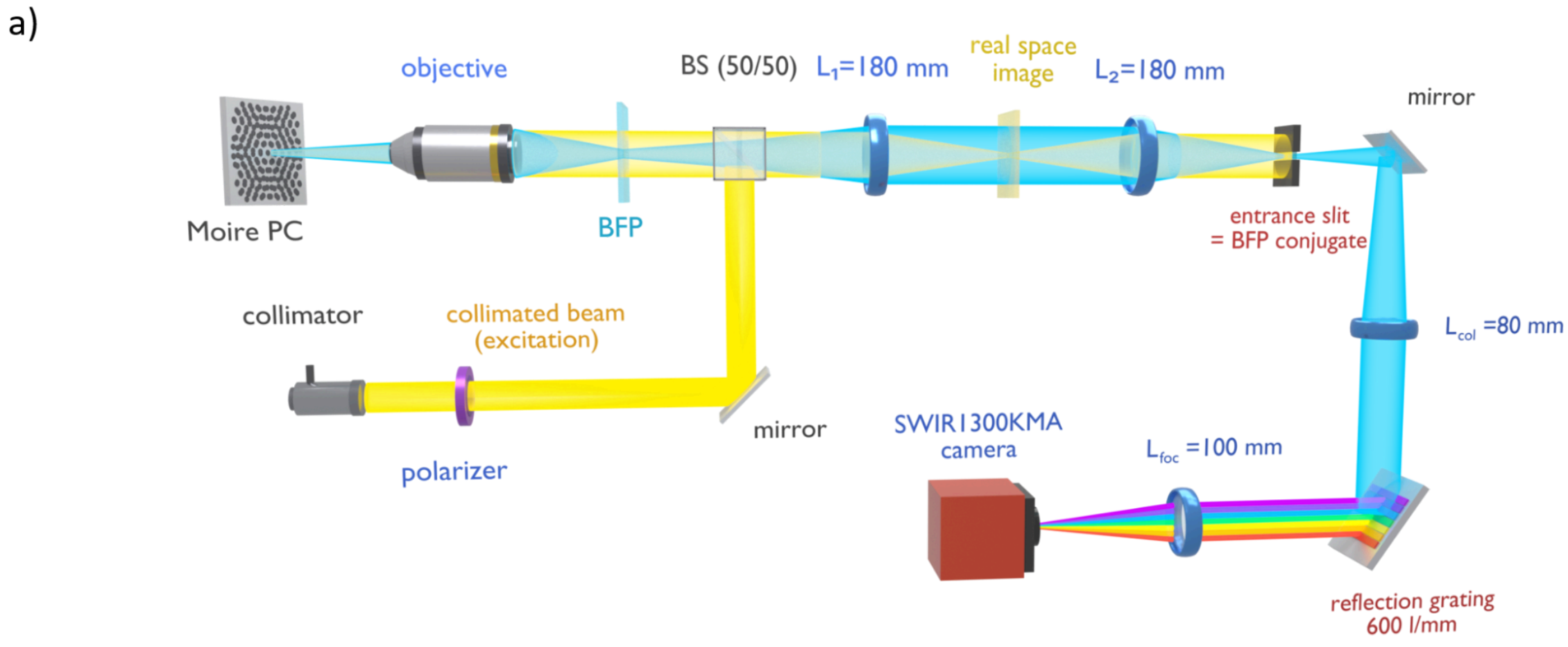

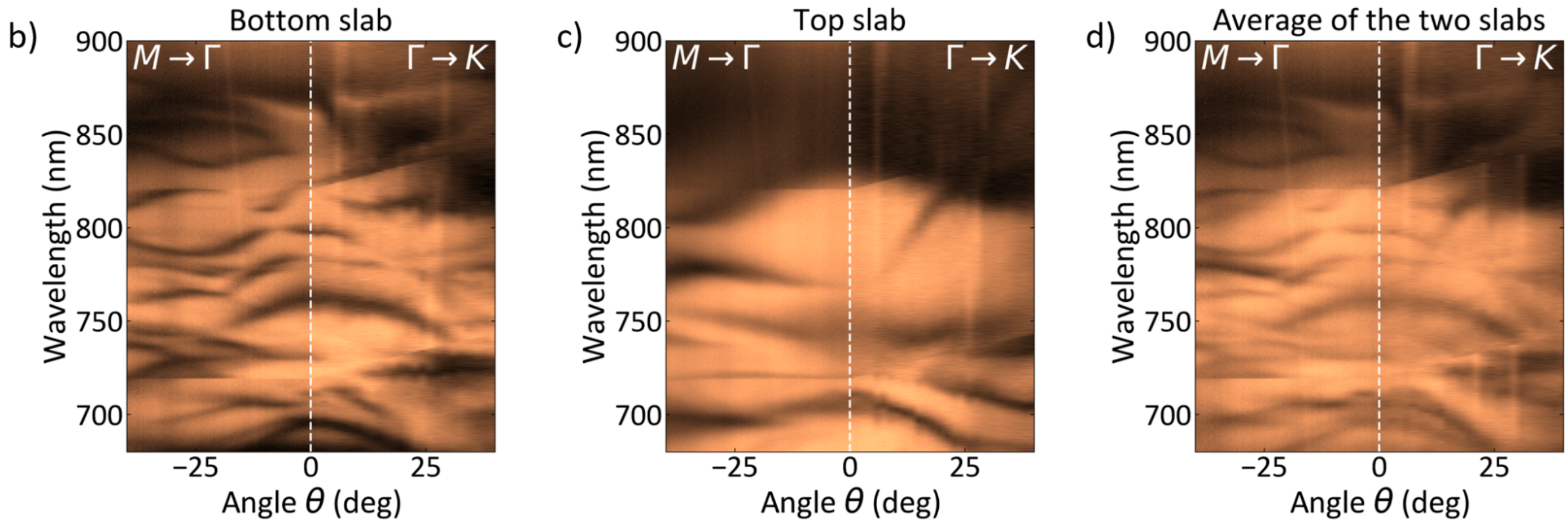

***Figure 3. Experimental setup for angle-resolved reflectivity measurements and optical band structures of the constituent PhC slabs**. a) Schematic of the back-focal-plane spectroscopy setup used to measure the optical band structure of the moiré PhCs. A collimated broadband beam is focused onto the sample through a microscope objective and the reflected signal is collected in a confocal configuration. The objective back focal plane (BFP), containing angular information, is relayed onto the entrance slit of a custom imaging spectrometer and recorded using a SWIR camera. Each acquisition records reflectivity as a function of wavelength and $\theta_y$ for a fixed value of $\theta_x$. The full three-dimensional angular dispersion is reconstructed by scanning $\theta_x$. b,c) Measured angle-resolved reflectivity spectra of the bottom and top PhC slabs, respectively, along the $\Gamma \rightarrow M$ and $\Gamma \rightarrow K$ directions. The vertical lines mark the $\Gamma$ points separating the $\Gamma \rightarrow M$ and $\Gamma \rightarrow K$ segments of the dispersion. d) Arithmetic average of the spectra shown in b and c, used as a reference for comparison with the moiré PhCs.*

We now proceed to measure the band structure of the moiré PhCs. Fig. 4a-c compares the measured and simulated angle-resolved reflectivity spectra of moiré PhCs with twist angles of 3.8°, while Fig. 4d-f shows the same results for angles of 8.4°. Relative to the single-slab spectra shown above in Fig. 3b-d, both moiré structures exhibit a richer optical response, characterized by the emergence of additional dispersive resonances. The main experimental features are qualitatively reproduced by numerical simulations, supporting their interpretation as moiré photonic bands arising from the coupling of optical modes supported by the twisted PhC slabs. Simulated spectra of the individual slabs are provided in the Supporting Information (Fig. S5 and S6).

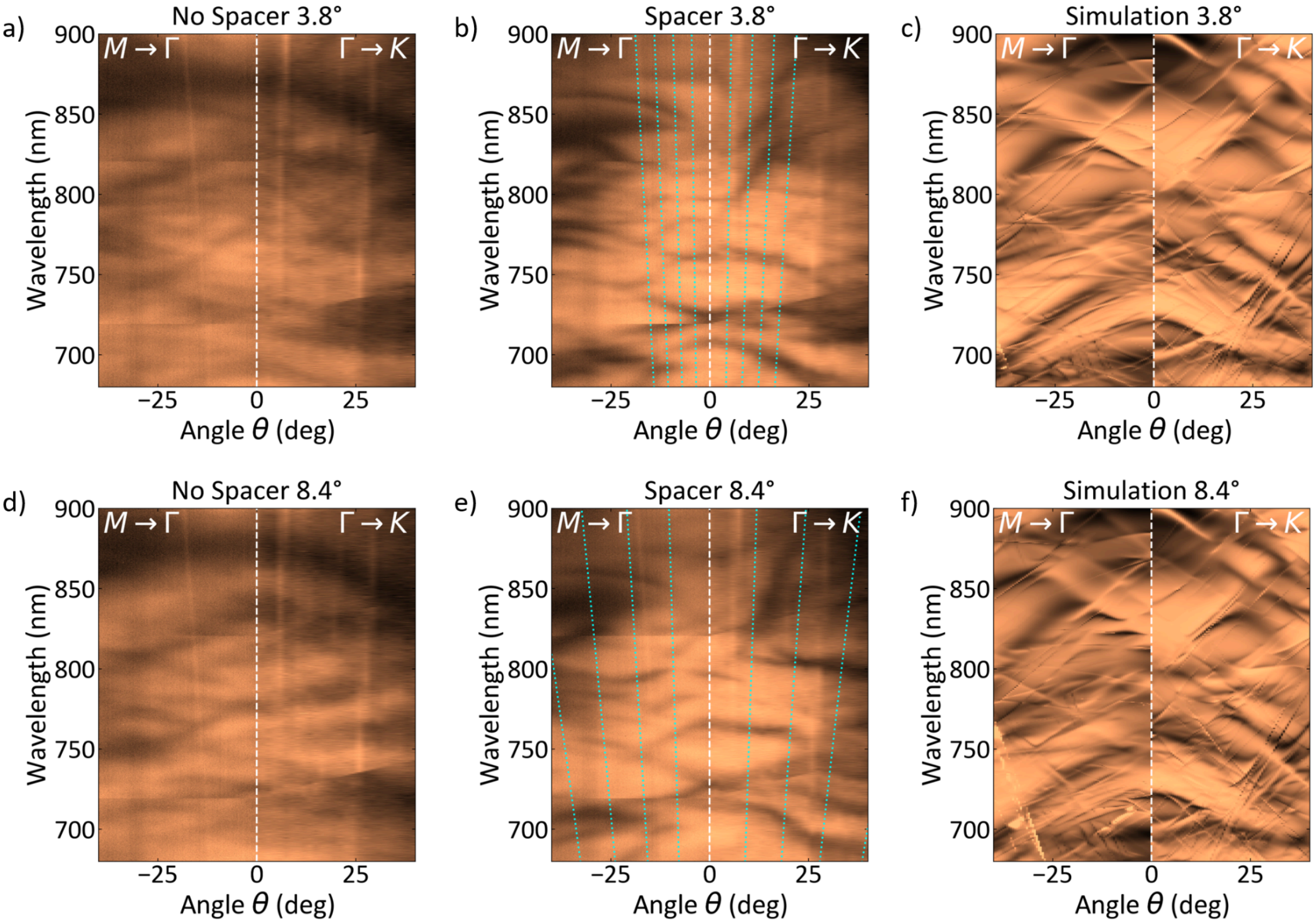

***Figure 4. Angle-resolved reflectivity spectra of twisted PhC slabs.*** *a) and d) Experimental angle-resolved reflectivity spectra of moiré PhC without an hBN spacer at twist angles of 3.8° and 8.4°, respectively. b) and e) Experimental angle-resolved reflectivity spectra of moiré PhCs with an hBN spacer at twist angles of 3.8° and 8.4°, respectively. Blue lines indicate the boundaries of the moiré Brillouin zones. c) and f) Corresponding simulated angle-resolved reflectivity spectra for the 3.8° and 8.4° structures. For clarity, only the angular range between -40° and 40° is displayed.*

The role of the interlayer spacing is highlighted by comparing the 3.8° and 8.4° moiré PhC with (Fig. 4a and d) and without an hBN spacer (Fig. 4b and e). Although both structures share the same moiré periodicity, their optical responses differ markedly. In the presence of the hBN spacer, multiple moiré bands remain spectrally distinguishable and can be directly resolved in the reflectivity measurements. By contrast, removing the spacer results in a dense spectrum of overlapping resonances that obscures the underlying band structure. This behaviour indicates that bringing the two PhC slabs into direct proximity substantially enhances the photonic inter-slab coupling, leading to stronger mode hybridization and an increased density of optical states. As a consequence, neighbouring resonances become less spectrally isolated, hindering the identification and selective addressing of individual moiré bands. These observations demonstrate that the optical response is governed not only by the presence of inter-slab coupling but also by its magnitude. The hBN spacer therefore plays a key role in establishing an intermediate coupling regime in which moiré-induced band formation remains pronounced while preserving spectrally distinct modes. Such a regime is advantageous not only for the direct observation of moiré photonic bands but also for applications that rely on selective coupling to individual photonic states.

Comparison of the 3.8° and 8.4° devices reveals an increased number of observable resonances at smaller twist angles, both experimentally and in simulation. The 3.8° moiré PhC exhibits a richer band structure than the 8.4° device, with additional dispersive features appearing throughout the measured spectral range. In particular, the 3.8° structure exhibits a larger number of resolvable resonances within ±15° of normal incidence, especially within the 790-830 nm spectral range, whereas fewer features are observed for the 8.4° device. This behaviour is consistent with the evolution of the moiré periodicity discussed above. As the twist angle decreases, the moiré lattice constant increases and the corresponding moiré Brillouin zone shrinks, leading to enhanced folding of the photonic bands into a reduced momentum-space volume. The calculated boundaries of the first moiré Brillouin zones, indicated by blue lines in Fig. 4b and e, move closer to normal incidence as the twist angle decreases, reflecting this reduction of the moiré Brillouin zone. Consequently, a larger number of folded states become visible within the measured dispersion maps, resulting in the richer dispersion patterns observed at smaller twist angles. Experiment and simulation exhibit the same trend, with additional resonances emerging at smaller twist angles, consistent with the formation of additional moiré photonic bands.

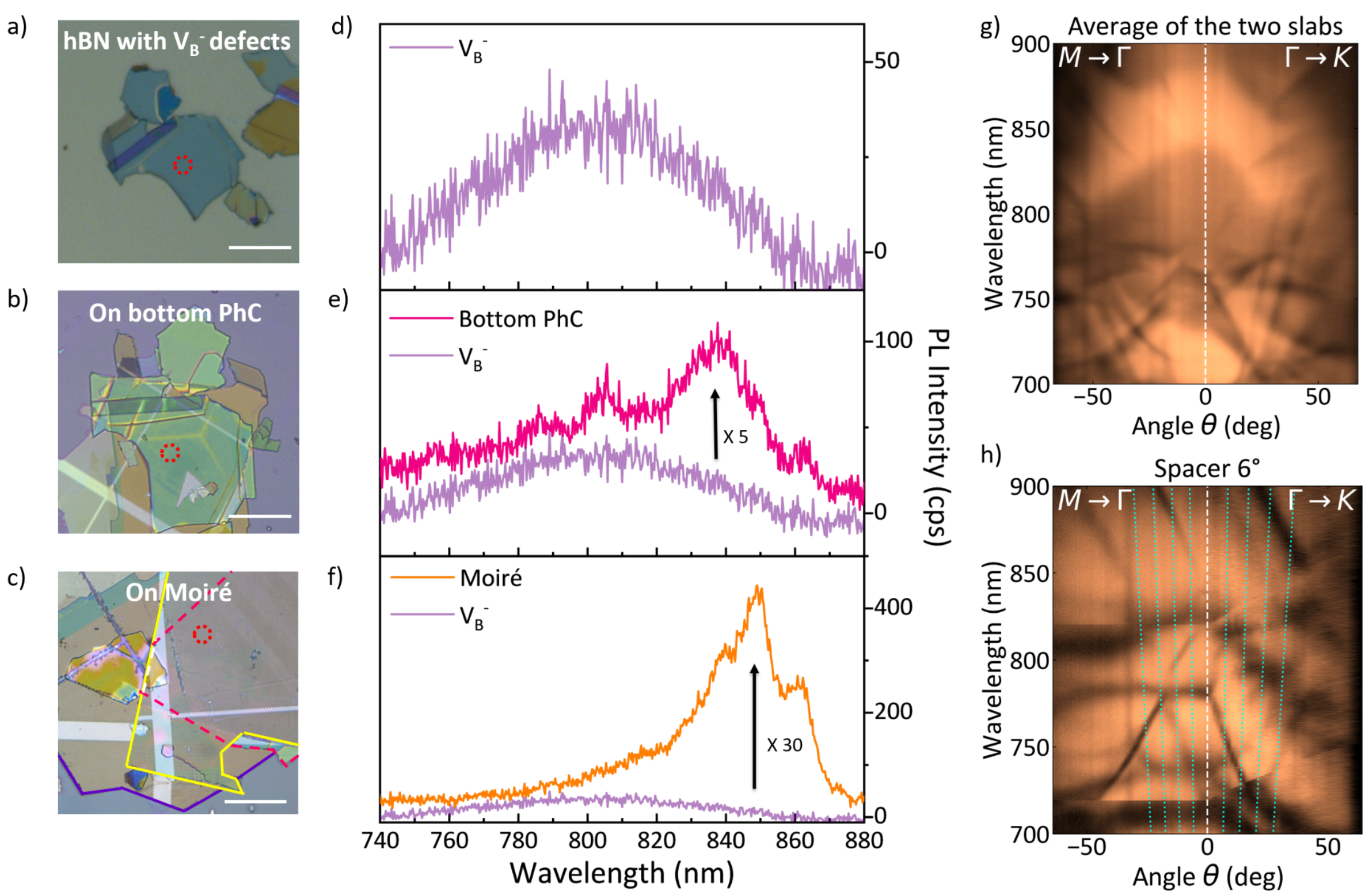

***Figure 5. PL enhancement of $V_B^-$ emission in single and twisted bilayer PhC slab** a-c) Optical micrographs of hBN flake with $V_B^-$ defects, the hBN flake on bottom PhC slab, and twisted bilayer PhC slabs, respectively. d-f) PL spectra of hBN flake with $V_B^-$ defects, after integrating with the bottom PhC slab, and moiré structure, respectively. g) and h) Angle-resolved reflectivity spectra of the constituent top and bottom PhC slabs and the moiré twisted bilayer PhC slabs, respectively.*

The increased density of photonic states associated with moiré band formation is expected to influence light-matter interactions. As a proof of concept, we integrate optically active defects in hBN with the moiré bilayer PhC slabs. In this geometry, a hBN flake with the boron vacancy ($V_B^-$) defects act as both an emissive channel and interlayer spacer to couple to the moiré cavity. Fig. 5a-c show the optical images of hBN flake with $V_B^-$ defects, the same hBN flake on bottom PhC slab, and after assembling bilayer PhC slabs at 6°, respectively.

Before assembly, the hBN flake exhibits a broad emission band centred around 800 nm, consistent with the characteristic emission of $V_B^-$ defects **(*38*, *39*)** (Fig. 5d). Note, the PL spectra are recorded from the same spot (marked as red dotted circles in Fig. 5a-c). A slight enhancement of the PL intensity with distinct peaks across $V_B^-$ emission from 770 to 870 nm is observed once the hBN is positioned on top of the bottom moiré PhC slab (Fig. 5e). These peaks are associated with the dispersive resonances from the bottom PhC slab (Fig. S8), suggesting coupling between $V_B^-$ emitters and the photonic modes supported by the bottom PhC slab. Remarkably, once the top slab is transferred and the moiré cavity is complete, the PL intensity is enhanced 30-fold compared to the bare $V_B^-$ emission. The largest increase occurs within the 825-875 nm spectral range, with a pronounced maximum near 850 nm, as shown in Fig. 5f.

To correlate PL enhancement with the moiré bands, we further performed angle-resolved reflectivity measurement of the moiré PhC slabs. Fig. 5g shows the measured optical band structure of the average top and bottom PhC slabs, while Fig. 5h shows the bands of the moiré PhC twisted at 6°. The constituent PhC slabs provide only a limited density of optical states within the $V_B^-$ emission range. In contrast, the moiré structure enables a dense manifold of additional photonic modes stemming from moiré-induced band folding. The largest increase in photonic band density occurs between 825 and 875 nm, coinciding with the spectral region exhibiting the strongest PL enhancement of the VB- emission. The spectral correlation between the enhanced emission and the reconstructed moiré band structure indicates that the moiré superlattice substantially modifies the photonic environment experienced by the $V_B^-$ centres. The increased overlap between the emitters and the optical modes of the assembled structure further reinforces this effect. These observations highlight the potential of moiré PhCs to manipulate light-matter interactions through twist-controlled photonic band engineering.

## DISCUSSION

The results presented here establish the twist degree of freedom as a versatile platform for engineering photonic band structures through the combined control of twist angle and inter-slab coupling. In contrast to conventional PhCs, whose optical properties are largely determined during fabrication, the moiré architecture introduces an additional control knob through the mechanical rotation of two PhC slabs. As demonstrated experimentally, reducing the twist angle increases the moiré periodicity and leads to a larger number of observable resonances, resulting in an increased density of optical states. These observations illustrate how geometrical control at the device level can be used to reshape the photonic band structure through moiré band formation and hybridization. A second and equally important control parameter is provided by the inter-slab spacing. The insertion of an hBN spacer between the two PhC slabs enables the interaction between optical modes supported by the two slabs to be tuned independently of the twist angle. The comparison between structures with and without an hBN spacer reveals that the magnitude of the inter-slab coupling strongly influences the visibility and spectral isolation of the resulting moiré photonic bands. While direct contact between the slabs produces a dense manifold of strongly overlapping resonances, the hBN spacer establishes an intermediate coupling regime in which moiré-induced band formation remains pronounced while preserving spectrally distinguishable optical modes. These observations demonstrate that the appearance of moiré photonic bands depends not only on the presence of inter-slab coupling but also on its magnitude.

Combined, the twist angle and inter-slab spacing enable a complementary control over the photonic band structure: the twist angle primarily governs the degree of band folding and the resulting density of optical states, whereas the inter-slab spacing regulates the strength of mode hybridization between the two PhC slabs.

To summarise, we have demonstrated moiré PhCs fabricated from vdW crystals, and experimentally observed the formation of moiré photonic bands through angle-resolved reflectivity measurements. The engineered moiré PhC resulted in a 30-fold enhancement from the embedded optical defects in the hBN spacer layer, establishing a direct connection between moiré photonic band formation and light-matter interactions. More broadly, these results demonstrate that twist-controlled photonic band engineering provides a powerful route for tailoring the optical environment experienced by solid-state emitters.
Future implementations could further expand the capabilities of this platform. For instance, combined with MEMS-based actuation schemes capable of dynamically controlling both the twist angle and inter-slab separation could provide access to continuously tunable moiré configurations (***40*, *41***). Such architectures could replace the fixed spacer layer with a tunable air gap, enhancing optical confinement while providing real-time control of the moiré band structure. Access to continuously tunable moiré configurations may enable the exploration of localized photonic states and flat-band regimes, opening new opportunities for dynamically engineering light-matter interactions in photonic moiré systems.

## METHODS

1. $WS_2$ photonic crystals fabrication

The $WS_2$ PhCs are fabricated on 1μm $SiO_2$/Si substrates purchased from Inseto. First, the substrate is cleaned by n-Butyl acetate (20 min), acetone (20 min) and isopropanol (IPA) (10 min) in an ultrasonic water bath and then blow-dried with nitrogen; after that, we treat the substrate in oxygen plasma to remove the residues and contaminants. During the cleaning procedure, $WS_2$ crystals from HQ graphene were mechanically exfoliated and then transferred on the clean substrate right away with 65 °C temperature to promote adhesion.

Second, the sample is prepared for Electron beam lithography (EBL). The spin coating procedure consists of two steps: first, spin coating the positive electron-beam resist CSAR-62 (AR-P-6200.13) at 4000 rpm for 30 s and baking on a hot plate at 180 °C for 2 mins; then spin coating a conductive layer of Electra 92 (AR-PC-5090.02) at 4000 rpm for 30s and baking on a hot plate at 100 °C for 1 min to mitigate charging effects that could reduce the patterning resolution during EBL. In the EBL step, the PhC design is patterned into the resist layer using an EBL machine (Raith Voyager) at 50 kV accelerating voltage and 1.21 nA beam current. To determine the optimum exposure dose, a dose test is typically performed prior to the fabrication of the actual sample layout. This entails patterning multiple copies of a test design, each with slightly different exposure doses, which are then checked with an SEM to find the optimum value.

After the patterning process, the sample is immersed in DI water for 1 min to remove the layer of Electra 92, then in xylene for 2 mins to dissolve the exposed areas of the positive resist CSAR 62, and finally in IPA to get rid of chemical residues prior to blow-drying. The patterned layer of resist covering the sample surface acts as a mask for the subsequent RIE, upon which the design is transferred into the $WS_2$ flake. Before reactive ion etching (RIE) our sample, we cleaned the chamber with Ar + $H_2$ and $O_2$ plasma. We used a combination of $CHF_3$ and $SF_6$ plasma for 40s with 0.14 mbar pressure and a DC bias of 135 V to provide a mixture of physical and chemical etching. After successful RIE, the residual resist film is removed by immersion in a hot 1165 resist remover (90 °C, 30 min) and hot acetone (90 °C, 30 min), followed by a rinse with IPA and several seconds of $O_2$ plasma ashing to remove the harder resist residues caused by RIE. Electron microscope images of the resulting samples were obtained using a scanning electron microscope (Thermo Fisher Phenom X). The thickness of the flakes and resulting holes (see Figure S2) were measured using atomic force microscope (Bruker Icon).

## 2. Assembly of twisted bilayer $WS_2$ photonic crystals

The twisted bilayer heterostructure (PhC/hBN/PhC) were assembled and twisted using a dry transfer method with a custom-built transfer setup. The objective lens was used to view the sample through the stamp, the manual xyz stage used for controlling the stamp and the heated sample stage that can be manually rotated as well as being moved in x and y directions. To connect the stamp to the xyz stage, a metal arm with a hole in it is used and is removable for attaching stamps and positioning substrates. For the stamp, a dome shaped polydimethylsiloxane (PDMS; SYLGARD™ 184 Silicone Elastomer, Dow, MI) stamp coated with a thin film of polyvinyl alcohol (PVA) polymer was used. The stamp was used to pick up one of the $WS_2$ PhCs at 60°C, followed by hBN flake. Then, the bottom $WS_2$ PhC was aligned at angle $\theta$ and release the stack (PhC/hBN) at 120 °C. Residual PVA polymer was removed by dissolving it in water at 40 °C for 2h. The final stack was dried by $N_2$ before measurement.

## 3. Computational modelling

The optical band structures of the moiré PhCs were calculated using RCWA4D, a rigorous coupled-wave analysis (RCWA) solver developed at Stanford University that enables the simulation of structures with two distinct periodicities and is therefore well suited for moiré photonic systems. The experimentally measured geometrical parameters, including lattice constant, slab thicknesses, inter-slab separation, and twist angle, were used as inputs to the simulations.

The optical response of the individual PhC slabs, presented in the Supplementary Information, was calculated using the Stanford Stratified Structure Solver (S4), an RCWA-based electromagnetic solver for periodic photonic structures. The simulated reflectivity spectra were obtained using the same geometrical parameters as those measured experimentally.

## 4. Back focal plane angle-resolved reflectivity measurement

Angle-resolved reflectivity measurements were performed using a back-focal-plane (BFP) imaging setup coupled to an imaging spectrometer. The collected signal is relayed through a 4f imaging system, with the first lens positioned one focal length from the objective BFP. This configuration forms a real-space image at the intermediate focal plane between the two lenses, where an adjustable pinhole is placed to define the probed area on the sample. The second lens then re-images the BFP onto the entrance slit of a custom imaging spectrometer.

The spectrometer consists of a relay-lens system, a diffraction grating, and a SWIR camera. The entrance slit selects a narrow range of reflection angles along the $\theta_x$ direction, while the full range of $\theta_y$ values permitted by the numerical aperture of the objective is spectrally resolved and recorded on the camera. Consequently, each acquisition provides the reflectivity as a function of wavelength and $\theta_y$ for a fixed value of $\theta_x$. Full mapping of the angular dispersion is enabled by mounting the entrance slit on a motorized translation stage, allowing systematic scanning of $\theta_x$ (Fig. S4). The diffraction grating is additionally mounted on a rotation stage to extend the accessible spectral range. For each slit position, the wavelength calibration is adjusted accordingly. Combining measurements acquired at different slit positions yields the complete angularly resolved band structure, as illustrated in the bottom panel of Figure 3a.

**ACKNOWLEDGEMENT**

The authors acknowledge financial support from the Australian Research Council (CE200100010, FT220100053, DP250100973, DP260102670) and the Air Force Office of Scientific Research (FA2386-25-1-4044). The authors also acknowledge the facilities as well as the scientific and technical assistance of the Sydney Nano Foundry, Core Research Facility at the University of Sydney, part of the NSW node of NCRIS-enabled Australian National Fabrication Facility.

**COMPETING INTERESTS**

The authors declare that they have no competing interests.